\documentclass[twocolumn, prl, amssymb,  aps, showpacs,preprintnumbers,
amsmath,showkeys,floatfix]{revtex4}

\usepackage{epstopdf}
\usepackage{graphics}
\usepackage{graphicx}
\usepackage{dcolumn}
\usepackage{bm}
\usepackage{longtable}
\usepackage{epsfig}
\usepackage{times}
\usepackage{url}
\usepackage{color}

\begin{document}
\title{
Proposed Entanglement of X-ray Nuclear Polaritons as a Potential Method for Probing Matter at the Subatomic Scale }

\author{Wen-Te \surname{Liao}}
\email{Liao@mpi-hd.mpg.de}

\author{Adriana \surname{P\'alffy}}
\email{Palffy@mpi-hd.mpg.de}

\affiliation{Max-Planck-Institut f\"ur Kernphysik, Saupfercheckweg 1, D-69117 Heidelberg, Germany}
\date{\today}
\begin{abstract}
A setup for generating the special superposition of a simultaneously forward- and backward-propagating collective excitation in a nuclear sample is studied. 
We show that by actively manipulating the scattering channels of single x-ray quanta with the help of a normal incidence x-ray mirror, a nuclear polariton which propagates in two opposite directions can be generated. The two counterpropagating polariton branches are entangled by a single x-ray photon. The quantum nature of the nuclear excitation entanglement gives rise to a subangstrom-wavelength standing wave excitation pattern that can be used as a flexible tool to probe matter dynamically on the subatomic scale.

\end{abstract}
\maketitle

Optical lasers have been at the heart of quantum physics for the last decades. The commissioning of the first x-ray free electron lasers (XFEL) \cite{slac,Sacla,Emma2010.NP,Gutt2012.PRL,Ishikawa2012.NP} and the developments of x-ray optics elements \cite{SchroerAPL,SchroerPRL,KangAPL, Shvydko2010,Shvydko2011,mimura2013,Osaka2013} bring into play higher photon  frequencies and promote the emerging field of x-ray quantum optics \cite{Adams2013}. Apart from a powerful imaging tool,  coherent x-ray light may also offer a solution for avoiding the diffraction limit bottleneck for compact  photonic devices \cite{Politi2008,Liao2012a}, and render possible the coherent control of transitions in  ions \cite{Buth2007,Young2010.N,Kanter2011.PRL,Zoltan2011.PRL,Rohringer2012} and nuclei \cite{Baldwin1997,Kocharovskaya1999,Buervenich2006, Liao2011, Piazza2012, Liao2013, Olga2013}. Furthermore, new  x-ray optical elements such as beam splitters \cite{Osaka2013} and mirrors \cite{Shvydko2010,Shvydko2011}  lay the foundation for 
developing quantum interferometers and controlling the quantum behavior of single x-ray photons. Apart from their potential in the field of quantum information \cite{Bennett1993, Lombardi2002}, the probing proficiency of single x-ray quanta would be an appreciated counterpart to traditional x-ray imaging techniques with intense x-ray beams \cite{xfel}.

In this Letter we present a coherent control scheme  that exploits the quantum properties of  a  single x-ray photon to generate nuclear polariton entanglement and a subangstrom-wavelength standing wave excitation pattern. The key for our setup is a  special superposition of a simultaneously forward- and backward-propagating collective excitation in a nuclear sample. The system comprising a single x-ray photon resonantly propagating through a sample of $^{57}\mathrm{Fe}$ M\"ossbauer nuclei forms a nuclear polariton \cite{Haas1988,Roehlsberger2004,Smirnov2005,Smirnov2007}.
We show how with the help of a normal incidence x-ray mirror \cite{Shvydko2011}, this polariton can be split in two entangled counterpropagating branches that share the same single photon. By actively distributing the single-photon wave packet in a controlled way, both the symmetric and the antisymmetric versions of  polariton entanglement can be achieved. This leads to the creation of a standing-wave nuclear excitation pattern   with subangstrom-wavelength in the absence of any x-ray  cavity or Bragg condition in the sample. In conjunction with phonon excitation, this standing wave can be used to dynamically probe the sample lattice using a single x-ray photon.

\begin{figure*}[ht]
\begin{center}
\vspace{-0.4cm}
  \includegraphics[width=1\textwidth]{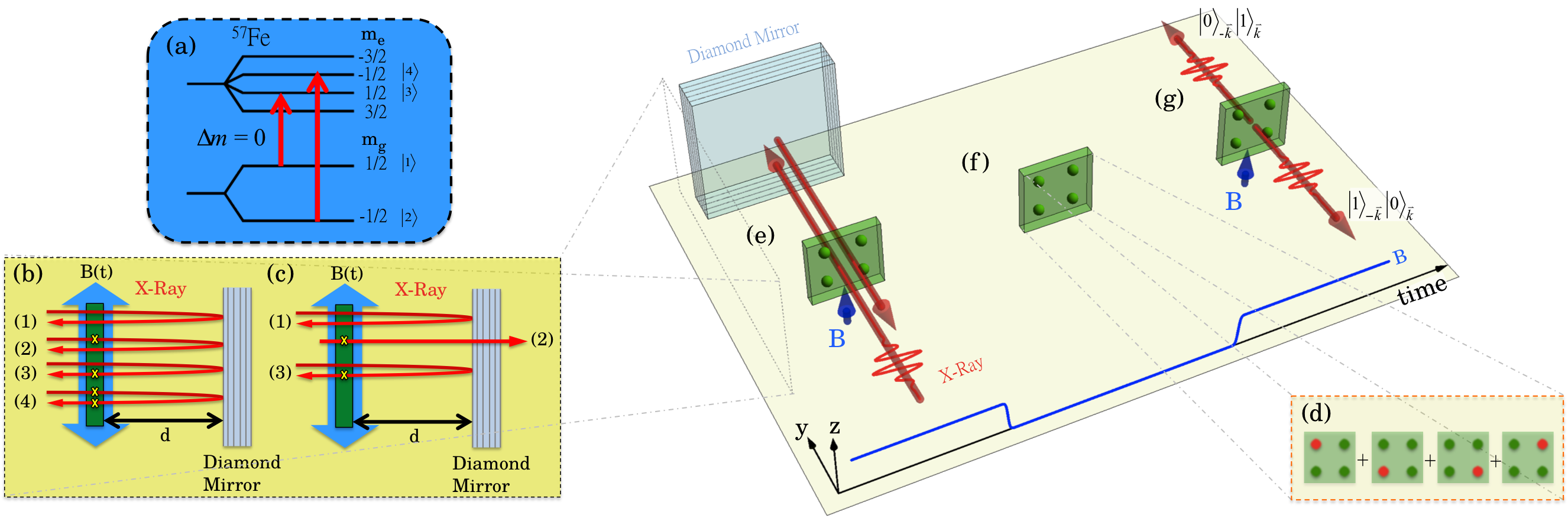}
  \caption{\label{fig1}
(Color online) (a) $^{57}$Fe ground and first excited state nuclear hyperfine levels. The $\Delta m=0$ transitions are driven by linearly polarized x rays.  
(b) NFS setup with a normal incidence x-ray diamond mirror placed at the distance $d$ from the $^{57}$Fe sample (green rectangle). The blue thick vertical arrow shows the applied magnetic field  {\bf B}(t). The possible scattering channels are depicted by the red thin arrows while yellow crosses illustrate the nuclear excitation. 
(c) The two scattering channels producing nuclear polariton entanglement. 
(d) Sketch of nuclear polariton. Red (green) spheres illustrate nuclei in the excited (ground) state.   
(e) Steps for achieving nuclear polariton entanglement. Green cuboid shows the $^{57}$Fe target and the red long arrows illustrate the incident, reflected, and emitted x rays. The short blue arrows depict the applied magnetic field {\bf B}(t).
The mirror reflection  causes an overlap of forward and backward single x-ray photon wave packets. 
(f) By switching off the hyperfine field {\bf B}(t), the single photon entanglement state is transferred to nuclear polariton entanglement. 
(g) The single-photon entanglement state $\vert \mathrm{SPE}\rangle$ is retrieved by turning on the 
{\bf B}(t) field. 
  }
\end{center}
\end{figure*}

The underlying physics for the creation and control of the nuclear polariton is the collective excitation of an ensemble of identical M\"ossbauer nuclei in solid-state targets, routinely observed in  nuclear forward scattering (NFS) experiments \cite{Roehlsberger2004}.
A monochromatic x-ray pulse with meV bandwidth tuned on the nuclear transition energy of 14.413 keV between the ground and the first excited state of  $^{57}$Fe shines perpendicular to the  solid-state target as shown in Figs.~\ref{fig1}(a)-(c). Our setup requires that synchrotron radiation or a low-intensity XFEL pulse excite at most one nucleus throughout the whole sample. In coherent scattering, the absence of nuclear recoil or spin flip does not allow identification of the nucleus involved in the scattering process. Thus, the nuclear polariton state is created
$\vert \mathrm{NP} \rangle=\frac{\alpha(t)}{\sqrt{N}}\sum_{\ell=1}^{N}\vert g\rangle\vert e_{\ell}\rangle e^{i\vec{k}\cdot \vec{r}_{\ell}}\otimes\vert 0 \rangle + \beta(t)\vert G \rangle\otimes\vert 1 \rangle_{\vec{k}}$.
The states denoted by  $\vert g\rangle\vert e_{\ell}\rangle$  are illustrated in Fig.~\ref{fig1}(d) referring to the situation when only the $\ell$th nucleus at position $\vec{r}_{\ell}$ is excited by the incident x ray with the wave number $\vec{k}$, whereas all other $(N-1)$ nuclei remain in their ground state. For the photon part, $\vert 0 \rangle$ denotes the zero-photon number state. The state $\vert G \rangle\otimes\vert 1 \rangle_{\vec{k}}$ depicts the case with  all nuclei  in the ground state $|G\rangle$ and one single $\vec{k}$-mode photon that propagates in the nuclear medium. As  further notation, $\alpha(t)$ and $\beta(t)$ are time-dependent coefficients. The dynamical energy exchange (multiple scattering) \cite{Hannon1999} between the states $\sum_{\ell=1}^{N}\vert g\rangle\vert e_{\ell}\rangle e^{i\vec{k}\cdot \vec{r}_{\ell}}\otimes\vert 0 \rangle$ and  $\vert G \rangle\otimes\vert 1 \rangle_{\vec{k}}$ \cite{Smirnov2005} is known as dynamical beat and exhibits remarkable properties, e.g.,  $\
vec{k}$ directionality, speed-up and coherent emission of single x-ray quanta \cite{Roehlsberger2004}. The collective nature of the excitation allows controlling the decay of the whole nuclear ensemble by external fields \cite{Shvydko1996,Palffy2009,Liao2012a,Liao2012b}.

In Fig.~\ref{fig1} we put forward a method of actively distributing the single photon wave packet in two opposite directions using a normal incidence x-ray mirror. Such near-100 $\%$ Bragg reflectivity mirrors made of diamond crystals have only recently been developed \cite{Shvydko2010,Shvydko2011}. The aim is to generate a nuclear polariton forward-backward momentum entanglement (NPE)
\begin{eqnarray}
\vert \mathrm{NPE} \rangle 
&=&\frac{\alpha(t)}{2\sqrt{\pi N}}\sum_{\ell=1}^{N}\vert g\rangle\vert e_{\ell}\rangle\left( e^{i\vec{k}\cdot \vec{r}_{\ell}}\pm e^{-i\vec{k}\cdot \vec{r}_{\ell}}\right)\otimes\vert 0 \rangle \nonumber\\
&+& \frac{\beta(t)}{\sqrt{2}}\vert G \rangle\otimes\left( |1\rangle_{-\vec{k}}|0\rangle_{\vec{k}}\pm |0\rangle_{-\vec{k}}|1\rangle_{\vec{k}}\right) \, .
\label{eq2}
\end{eqnarray}
The spatial part of $\vert \mathrm{NPE} \rangle$ shows that  the coherently emitted single photon wave packet  propagates in both forward  ($\vec{k}$) and backward ($-\vec{k}$) directions. However, due to the particle nature of a single photon, the signal can be registered either at the forward or at the backward detector. A single photon forward-backward entanglement state (SPE) is thus generated
$\vert \mathrm{SPE}\rangle=(|1\rangle_{-\vec{k}}|0\rangle_{\vec{k}}\pm |0\rangle_{-\vec{k}}|1\rangle_{\vec{k}})/\sqrt{2}$. Single-photon entanglement \cite{Enk2005} refers to entanglement of two spatially distinguishable field modes by a single photon and lies at the heart of quantum networks \cite{Kimble2008,Morin2013}. In our setup, $|1\rangle_{-\vec{k}}|0\rangle_{\vec{k}}$ is associated with registering the photon at the backward detector and 
 $|0\rangle_{-\vec{k}}|1\rangle_{\vec{k}}$ at the forward detector, in either case with no photon counts at the detector placed in the respective opposite direction.  The single excitation is an important feature for our setup, since incoherent multiple excitation  would change both the spatial 
and the photon part of $\vert \mathrm{NPE} \rangle$.

We start by investigating the system depicted in Fig.~\ref{fig1}(b). Scattering in a typical NFS setup involving one target proceeds by creating either one or no excitation. Two more channels are introduced by the mirror reflection, leading to four possibilities: (1) no scattering occurs, (2) one excitation is produced and the single resonant photon is scattered in the forward direction, (3) a single excitation occurs only after the mirror reflection and the scattered photon thus proceeds in the backward direction, and (4) on both passages through the sample the x-ray pulse creates one excitation. Because of the reflection, the effective resonant thickness \cite{Crisp1970, Shvydko1998, Shvydko1999N} of the target for the backward time spectra will be doubled, just as in the case of a scattering of two identical samples \cite{Smirnov2005}. 
For generating $\vert \mathrm{NPE} \rangle$ and $\vert \mathrm{SPE}\rangle$, the setup must be modified to fulfill the 
 following conditions: (i)  channel (2) should pass through the mirror without being reflected, (ii) channel (4) should be suppressed such that the occurrence of forward and backward scattering events have equal probability, and (iii) the relative phase between channels (2) and (3) 
needs to be tunable such that the single photon wave packets of the two channels will form the $\vert \mathrm{SPE}\rangle$ state. 

Figure~\ref{fig1}(c) is an illustration of the situation described above.
By reflecting only the incident x-ray pulse of channel (1) after it passed through the target, the conditions (i) and (ii) can be fulfilled simultaneously.
Because of the unique temporal structure of NFS, the prompt signal of channel (1)  can be resolved before the emergence of channels (2), (3), and (4) \cite{Ruby1974, Shenoy2007, Roehlsberger2004}. Thus, if the mirror reflection is quickly disabled after the incidence of the prompt x-ray pulse, 
the portion of signal needed for triggering channel (3) is temporally selected and channel (2) is allowed to pass. 
To achieve condition (iii), one could rely on either changing the distance $d$ between the nuclear target and mirror
or on the magnetic phase modulation \cite{Liao2012a} to control the relative phase between the forward channel (2) and the backward channel (3). In the following, we combine both methods and provide a  mechanics-free 
solution to modulate the relative phase between the two nuclear components of $\vert \mathrm{NPE} \rangle$.

To describe the dynamics of our  system depicted in Fig.~\ref{fig1}, the Maxwell-Bloch equations \cite{Crisp1970, Shvydko1999N, Scully2006, Palffy2008, Liao2012a} for the  density matrix $\widehat{\rho}$, 
$\partial_{t}\widehat{\rho} = \frac{1}{i\hbar}\left[ \widehat{H},\widehat{\rho}\right]+\widehat{\rho}_{s}$ and $\frac{1}{c}\partial_{t}\Omega+\partial_{y}\Omega =i\eta\left(a_{31}\rho_{31}+a_{42}\rho_{42}\right)$ are used. We adopt the forward-backward decomposition \cite{lin2009}, $\rho_{31}\rightarrow f_{31}e^{iky}+b_{31}e^{-iky}$, $\rho_{42}\rightarrow f_{42}e^{iky}+b_{42}e^{-iky}$ and $\Omega\rightarrow \Omega_{F}e^{iky}+\Omega_{B}e^{-iky}$. Here, 
$f_{eg}$ and $b_{eg}$ for $g\in \{1,2\}$ and $e\in \{3,4\}$  are the forward and the backward coherences, respectively, for the nuclear wave function $|\psi\rangle= A_{1}|1\rangle+A_{2}|2\rangle+A_{3}|3\rangle+A_{4}|4\rangle$. The ket vectors are the eigenvectors of the two ground and two excited states hyperfine levels with $m_g=-1/2$, $m_g=1/2$, $m_e=-1/2$ and $m_e=1/2$, respectively, as shown in Fig.~\ref{fig1}(a).
  The Maxwell-Bloch equations in the linear region, i.e., $\vert\Omega_{F}\vert \ll \Gamma$ and $\vert\Omega_{B}\vert \ll \Gamma$,
then read 
\begin{eqnarray}
&&
\partial_{t}f(b)_{31} = -\left( \frac{\Gamma}{2}+i\Delta_{\bf B}\right) f(b)_{31}+i\frac{a}{4}\Omega_{F(B)}\, ,
\nonumber\\
&&
\partial_{t}f(b)_{42} = -\left( \frac{\Gamma}{2}-i\Delta_{\bf B}\right) f(b)_{42}+i\frac{a}{4}\Omega_{F(B)}\, ,
\label{eq4}
\end{eqnarray}
\begin{eqnarray}
&&
\frac{1}{c}\partial_{t}\Omega_{F}+\partial_{y}\Omega_{F}=i\eta a\left(f_{31}+f_{42}\right)\, ,
\nonumber\\
&&
\frac{1}{c}\partial_{t}\Omega_{B}-\partial_{y}\Omega_{B}=i\eta a\left(b_{31}+b_{42}\right)\, .
\label{eq5}
\end{eqnarray}
In Eqs.~(\ref{eq4}), $\Delta_{\bf B}$ denotes the Zeeman energy shift of the nuclear transitions  proportional to the magnetic field {\bf B}.   Furthermore,  $a_{31}=a_{42}=a=\sqrt{2/3}$ is the corresponding Clebsch-Gordan coefficient \cite{Shvydko1998,Palffy2008,Liao2012a} for the  $\Delta m=0$ transitions.  The parameter $\eta$ is defined as $\eta=\frac{6\Gamma}{L}\xi$, where $\Gamma=1/141.1$ GHz is the $\gamma$-decay rate of excited 
states, $\xi$ represents the effective resonant thickness \cite{Crisp1970, Shvydko1998, Shvydko1999N} and $L=10$ $\mu$m  the thickness of the target, respectively. Further notations are $c$ the speed of light and $\Omega_{F(B)}$ for the forward (backward) Rabi frequency which is proportional to the electric field $\vec{E}$ of the x-ray pulse  \cite{Scully2006, Palffy2008}. The forward scattered x rays are treated as incident field for the backward scattering, i.e., $\Omega_{B}(t,L)=-\sqrt{R}\Omega_{F}(t,L)$, where $R=0.99$ denotes the mirror reflectivity \cite{Shvydko2011}, and the additional minus sign originates from the mirror reflection.

The source terms in the right-hand side of Eq.~(\ref{eq5}) are proportional to $\cos\left(\Delta_{\bf B}t\right)$ since the coherences in Eq.~(\ref{eq4}) evolve as $\exp({\pm i\Delta_{\bf B}t})$. Their behavior can therefore be manipulated by dynamically changing the external hyperfine magnetic field {\bf B}(t). For instance, a NFS single photon wave packet can be modulated with a phase shift of $\pi$ by inverting {\bf B}(t) at the time instants when $\Delta_{\bf B}t=\frac{n}{2}\pi$ with $n$ odd \cite{Liao2012a}. Also, the coherent emission related to  the nuclear polariton can be suppressed by switching off \cite{Liao2012a} or rotating \cite{Shvydko1996} {\bf B}(t) at the same time instants. We use the coherent photon storage \cite{Liao2012a} to design the required  steps for producing $\vert \mathrm{NPE} \rangle$ and $\vert \mathrm{SPE}\rangle$ as illustrated in Figs.~\ref{fig1}(e)-(g). 
First, an x-ray diamond mirror \cite{Shvydko2011} is placed for a short time $t_d$ behind the $^{57}$Fe target at the distance of ${ d}=c\pi/\left( 2\Delta_{\bf B}\right)$. This causes a time delay $\tau$ between the backward and forward signals such that $\tau=2 d/c=\pi/\Delta_{\bf B}$. For $t<\tau$, only the forward signal is present, allowing phase modulation of channel (2) only  instead of both (2) and (3) at around $t=\tau/2$ by inverting {\bf B}(t) \cite{Liao2012a}. After the incident pulse was reflected by the mirror, it impinges on the $^{57}$Fe target and 
reaches the backward detector. At $t_d$ the reflection is disabled such that channel (2) spatially remains in the forward direction. Now we switch off {\bf B}(t) at $t=3\pi/\left( 2\Delta_{\bf B}\right)$ to suppress the coherent $\gamma$ decay \cite{Liao2012a} in both directions, as shown in Fig.~\ref{fig1}(f). This projects the nuclear polariton $\vert \mathrm{NPE} \rangle$ to its matter part and neither detector  records any signal.  After turning the magnetic field  {\bf B}(t) on, either the backward  or the  forward detector may 
register a single photon, i.e., $\vert \mathrm{SPE}\rangle$ is generated.

\begin{figure}[b]
\vspace{-0.4cm}
  \includegraphics[width=0.45\textwidth]{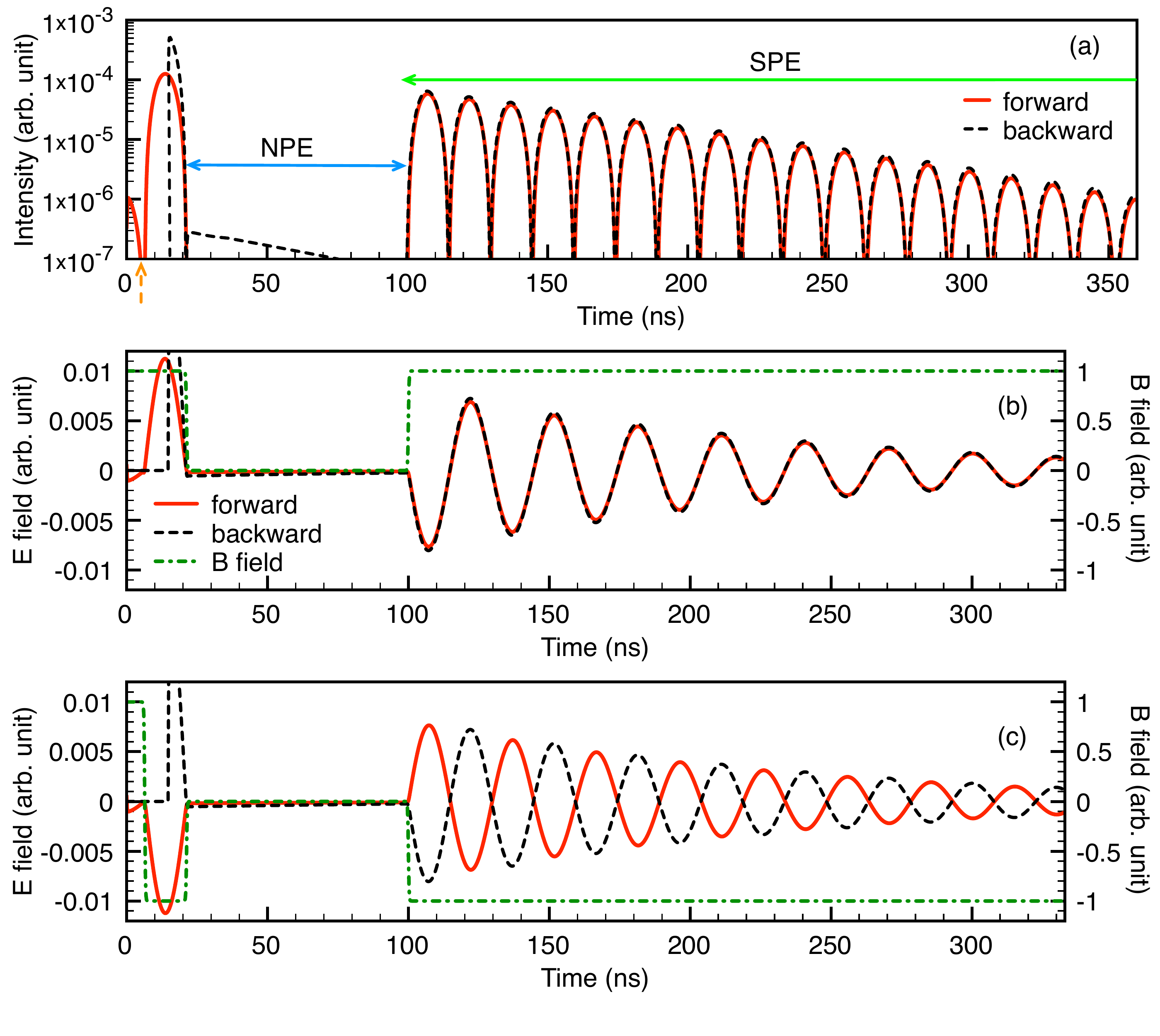}
  \caption{\label{fig2} (Color online) Time spectra  of coherently emitted single photons from a $^{57}$Fe nuclear target in the forward (red solid line) and backward (black dashed line) directions. (a) Scattered field intensity. The vertical dashed arrow indicates the time $t_d=$7.4 ns when the mirror reflection is disabled. 
(b), (c) Electric field amplitudes for different applied magnetic fields {\bf B}(t) (green dash-dotted lines). 
 (b) Production of the symmetric entanglement state $(|1\rangle_{-\vec{k}}|0\rangle_{\vec{k}}+|0\rangle_{-\vec{k}}|1\rangle_{\vec{k}})/\sqrt{2}$.
(c) Generation of the  antisymmetric entanglement state $(|1\rangle_{-\vec{k}}|0\rangle_{\vec{k}}-|0\rangle_{-\vec{k}}|1\rangle_{\vec{k}})/\sqrt{2}$ with an inversion of {\bf B}(t) at around 7.4 ns. 
}
\end{figure}

Our numerical results are demonstrated in Fig.~\ref{fig2} with $\Delta_{\bf B}=30\Gamma$ and $\xi=1$. A time delay around 15 ns between the two counterpropagating  signals  is introduced by the choice of $d$ for all cases.
In Fig.~\ref{fig2}(a), the forward and backward signal intensities are proportional to $\vert\Omega_{F}(t,L)\vert^{2}$ and $\vert\Omega_{B}(t,0)\vert^{2}$, respectively. For both forward and backward time spectra, the effective resonant thicknesses $\xi$ of the target are the same, due to our selective mirror reflection. Also, due to the high mirror reflectivity $R=0.99$, the residual forward signal intensity measured behind the mirror for $t <$ 7.4 ns is two orders of magnitude lower than that without reflection measured at later times. The applied magnetic field {\bf B}(t) is switched off at 22 ns and on at 100 ns. During this hyperfine-interaction-free period, the coherent decay of $\vert \mathrm{NPE} \rangle$ is significantly suppressed due to the destructive interference of two coherences  $f_{31}$ and $f_{42}$ ($b_{31}$ and $b_{42}$) for the forward (backward) field. Therefore, as indicated by the arrow in Fig.~\ref{fig2}(a), a time window with only the pure matter part of $\vert \mathrm{NPE} \rangle$ 
is built. By switching on the magnetic field at $t=100$ ns, the photon part of the polariton is retrieved and we can see the signature of the $\vert \mathrm{SPE}\rangle$ state. 
 The  equal probability of the two components of the state $\vert \mathrm{SPE}\rangle$ is achieved with an  
 adequate choice of  effective thickness $\xi$ and reflectivity $R$. For a small effective thickness, 
the signal is observed to experience a decay which can be roughly modeled by $\exp(-\pi\xi\Gamma/\Delta_{\bf B})$. The effect of the  mirror reflectivity on the backward signal with $R/\exp(-\pi\xi\Gamma/\Delta_{\bf B})\sim 1$ then renders the two signals almost identical for $t >$ 100 ns.

For complete control over the entangled state, we also address here how to switch the  identities of $\vert \mathrm{NPE} \rangle$ and $\vert \mathrm{SPE}\rangle$ between their symmetric and antisymmetric versions. The relative phase between the two components of the entangled state should then be modulated. Because of the mirror reflection, the forward and backward fields are  out of phase if the distance ${d}\ll 2c\pi/\Delta_{\bf B}$. Here, the purpose of the time delay of 15 ns is to make the two fields in phase as shown in Fig.~\ref{fig2}(b). Thus, without any magnetic phase modulation \cite{Liao2012a}, the symmetric $\vert \mathrm{NPE} \rangle$ and $\vert \mathrm{SPE} \rangle$ are automatically generated for $t >$ 22 ns. To produce the antisymmetric entangled state, we invert {\bf B}(t) at around 7.4 ns such that only the phase of the forward field is modulated with a shift of $\pi$. This antisymmetric $\vert \mathrm{SPE}\rangle$ is illustrated in Fig.~\ref{fig2}(c) for $t >$ 100 ns. The spatial part 
of the symmetric (antisymmetric) matter part of $\vert \mathrm{NPE} \rangle$ is a cosine (sine) standing wave that modulates the nuclear excitation pattern on a subangstrom scale. So far, x-ray standing waves involve the Bragg condition in crystals \cite{Woodruff} or the Borrmann  \cite{Batterman1964} or suppression  \cite{Afanas1965,Burck1978,Smirnov1980} effects. Furthermore, with the lack of 
x-ray cavities for normal incidence, they could so far only  be achieved for the first transversal mode of thin film x-ray cavities \cite{Roehlsberger2004,Roehlsberger2010,Roehlsberger2012,Evers2013}. 

Our  setup provides  robust and tunable single-photon entanglement. 
The emitted photon is at all times in an entangled state of the two counterpropagating field modes.  This is  a major improvement in comparison to a previous proposal to generate single-photon polarization entanglement \cite{Palffy2009}, where the two entangled field modes were temporally separated by $\approx$ 200 ns, making any experimental application difficult. Furthermore, the phase between the two entangled field modes can, in our scheme, be controlled via the magnetic field. An x-ray Mach-Zehnder interferometer setup \cite{Tamasaku2002,Shvydko2004} can be used to test both the generated entanglement (for instance, by  demonstrating the violation of a Bell inequality for a single photon \cite{LeeKim}) 
and the phase relation between the two branches in Figs.~\ref{fig2}(b) and (c). The $\vert \mathrm{SPE} \rangle$ components  forming the two interferometer arms are recombined on a beam splitter.  
Entanglement is then demonstrated by detection of target position or {\bf B}(t)-dependent interfered signals at the output ports \cite{Agarwal2012}.
Because of the high energy and momentum of the light quantum, our single-photon entanglement may  eventually be used to create macroscopical quantum superpositions for quantum measurement and decoherence studies \cite{Haroche1998}.

The quantum nature of $\vert \mathrm{NPE} \rangle$ creates a standing wave excitation pattern with subangstrom
wavelength in the nuclear sample. In contrast to the twin waves of the Borrmann \cite{Batterman1964} or suppression effects \cite{Afanas1965,Burck1978,Smirnov1980} or to the x-ray standing wave technique \cite{Woodruff} based on Bragg diffraction, the nuclear polariton is not limited by the crystal structure or by strict geometrical conditions. Its spatial pattern makes the nuclear excitation and the $\vert \mathrm{SPE} \rangle$ state sensitive to the nuclear position in the sample regardless of the sample structure.  In conjunction with optical-laser-driven \cite{OlgaHI} phonon excitation \cite{Sturhahn}, nuclear pump-probe experiments  or dynamical material imaging with resolution on the atomic scale can be envisaged.  The implementation of our setup  is related to two experimental challenges: the  fast disabling of the x-ray mirror reflection and the control over the tens-of-Tesla strong hyperfine magnetic field. The mirror reflection disabling can be achieved, for instance, by using a strong x-ray  focus together with a sub-ns piezoelectric Pb(Zr, Ti)O$_{3}$ (PZT) x-ray switch \cite{Grigoriev2006} to rapidly move the mirror in and out of the x-ray beam line. Alternatively, by taking advantage of the $\mu$rad sensitivity of Bragg reflection to the incidence angle, a fast rotation of the mirror \cite{Roehlsberger2000} can make the diamond crystal shift from reflection to transparency. Concerning the magnetic field control, the results presented here consider the situation when the hyperfine magnetic field can be switched off and on as discussed in Ref.~\cite{Liao2012a}. However, we expect that the nuclear excitation storage using the already demonstrated fast {\bf B}(t) rotation by means of  a 10 G external magnetic field \cite{Shvydko1996}  can also be utilized to produce the symmetric and antisymmetric $\vert \mathrm{NPE}\rangle$ and $\vert \mathrm{SPE}\rangle$. This is due to the fact that the relative phase between the two components of the  entanglement state are modulated 
before storage, and the phase evolution of the forward and backward components during the period of storage  are synchronous.

\bibliographystyle{apsrev}
\bibliography{NFSBS2A}
\end{document}